\documentclass[
    ,final            % use final for the camera ready runs
  ]
  {aipproc}

\layoutstyle{6x9}

\usepackage{amsmath}
\usepackage{amsfonts}
\usepackage{amssymb}

\newcommand{\ie}{\emph{i.e.}}

\begin{document}

\title{Simplicial Aspects of String Dualities}

\author{   Mauro Carfora, 
           Claudio Dappiaggi, 
           \underline{Valeria Gili}
           \footnote{contributing author}
         }{
  address={Dipartimento di Fisica Nucleare e Teorica,\\
  Universit\`a degli Studi di Pavia,\\
  and\\
  Istituto Nazionale di Fisica Nucleare, Sezione di Pavia,\\
  via A. Bassi 6, I-27100 Pavia, Italy}
}
\begin{abstract}
We will show how the study of randomly triangulated
surfaces merges with the study of open/closed string dualities. In
particular we will discuss the Conformal Field Theory which arises in
the open string sector and its implications.
\end{abstract}

\maketitle
%%%%%%%%%%%%%%%%%%%%%%%%%%%%%%%%%%%%%%%%%%%%
%% MAINMATTER
%%%%%%%%%%%%%%%%%%%%%%%%%%%%%%%%%%%%%%%%%%%%

String dualities provides a powerful tool to study IR properties of a
Quantum Field Theory by means of UV techniques proper of String
Theory. This approach had led to the formulation of AdS/CFT
correspondence but, at mathematical level it has been explained
only in a topological setting by Gopakumar and Vafa in
\cite{Gop-Vafa} and more recently by Gaiotto and Rastelli in \cite{Gai-Ras}.
In such a framework, a paradigmatical result has
been established by Gopakumar in \cite{Gop1,Gop2}. Starting
from a Schwinger 
parameterization of the free gauge field correlators of an
$\mathcal{N}=4$ SYM $SU(N)$ 
gauge theory, he exploits an analogy with
electrical networks which allows him to sum over internal loop
momenta to obtain a skeleton graph with a number of
vertices equal to the number of holes of the free open theory and
which shows the basic connectivity of the original correlator. By performing
a change of variables into the Schwinger parameter spaces, he is able
to fill the holes and obtain a closed AdS tree diagram.
In this context it is important to stress that planar graphs
with different connectivities give rise to different skeleton
diagrams, and all these different skeleton contributions need to be
summed over to obtain the closed string dual of a single open free
field diagram. Moreover, it is important to recognize that all these
structures are in one-to-one correspondence with the moduli space of a
sphere with $n$ holes, moduli space which arises as a natural structure in the
large N limit framework, proper of gauge/gravity correspondence.

Motivated by the ubiquitous role that simplicial methods play in the
above result, we have recently introduced a geometrical framework \cite{Mauro1} in
which it is 
possible to implement new examples of open/closed string dualities. Our
approach is based on a careful use of uniformization theory for
triangulated surfaces carrying curvature degrees of freedom. 

In order to show how this uniformization arises,
let us consider the dual polytope
associated with a Random Regge Triangulation Triangulation
\cite{Mauro3} $|T_l| \rightarrow M$ of a
Riemannian manifold $M$. 
Using the  
properties of Jenkins-Strebel quadratic differentials
\cite{Mauro1} it is possible to decorate the neighborhood of each
curvature supporting 
vertex with a punctured disk uniformized by a conical metric
\begin{equation*}
  ds_{(k)}^{2}\doteq 
  \frac{\lbrack L(k)]^{2}}{4\pi ^{2}}|\zeta (k)|^{-2(
    \frac{\varepsilon (k)}{2\pi })}|d\zeta (k)|^{2}.
\end{equation*}

Alternatively, we can blow up every such a cone into a corresponding
finite 
cylindrical end, by introducing 
a finite annulus 
$\Delta_{\varepsilon (k)}^\ast\doteq 
\left\{ 
\zeta(k) \in \mathbb{C} | \quad 
\exp{-\frac{2\pi }{2\pi -\varepsilon(k)}}
\leq |\zeta (k)|\leq 1
\right\}$
endowed with the cylindrical metric:
\begin{equation*}
  |\phi (k)|\doteq \frac{\lbrack L(k)]^{2}}{4\pi ^{2}}|
  \zeta (k)|^{-2}|d\zeta(k)|^{2} 
\end{equation*}

It is important to stress the different role that the deficit angle
plays in such two unformizations.   
In the ``closed'' uniformization the deficit angles $\varepsilon(k)$
plays the usual role of localized curvature degrees of freedom and,
together with the perimeter of the polytopal cells, provide the
geometrical information of the underlying triangulation.
Conversely, in the ``open'' uniformization, the deficit angle associated
with the $k$-th  polytope cell defines the geometric moduli
of the $k$-th cylindrical end. As a matter of fact each annulus can be 
mapped into a cylinder of circumpherence $L(k)$ and height $\frac{L(k)}{2 \pi -
  \varepsilon(k)}$, thus  
$\frac{1}{2 \pi - \varepsilon(k)}$ is the geometrical moduli
of the cylinder. 
This shows how the uniformization process works quite differently 
 from the one used in Kontsevich-Witten models, 
in which the whole punctured disk
is uniformized with a cylindrical metric. In this case the disk can
be mapped into a semi-infinite cylinder, no role is played by the
deficit angle and the model is topological;  conversely, in our case,
we are able to deal with a non topological theory.

In the closed sector both the coupling of the geometry of the
triangulation with $D$ bosonic fields and the quantization of the
theory can be performed under the paradigm of
critical field theory. However, in order to discuss Polyakov string theory
directly over the dual open Riemann surface so defined, we have to deal with a 
Boundary Conformal Field Theory (BCFT) defined over each cylindrical end.
The unwrapping of the cones into finite cylinders suggest to compactify
each field defined on the $k$-th cylindrical end along a circle of radius 
$\frac{R(k)}{L(k)}$: 
\begin{equation*}
  X^{\alpha }(k)
  \xrightarrow{\vartheta (k)\rightarrow \vartheta (k)+2\pi}
  X^{\alpha }(k) + 2\pi \nu^{\alpha} (k)\frac{R^{\alpha}(k)}{L(k)}
  \qquad
  \nu (k)\in \mathbb{Z}
\end{equation*}  

Under these assumptions, it is possible to quantize the theory and to
compute the quantum amplitude
over each cylindrical end: writing it as an amplitude between an
initial and final state, we can extract suitable boundary
states which arise as a generalization of the states introduced by
Langlands in \cite{Langlands}. 
As they stand, these boundary state do not preserve neither the
conformal symmetry nor the  $U(1)_L \times U(1)_R$ symmetry 
generated by the cylindrical geometry.
It is then necessary to impose on them suitable gluing
conditions relating the holomorphic and anti-holomorphic generators on
the boundary. These restrictions generate the usual families of
Neumann and Dirichlet boundary states.

Within this framework, the next step in the quantization of the theory is to define the
correct interaction of the $N_0$ copies of the cylindrical CFT on the ribbon
graph associated with the underlying Regge Polytope. 
This can be achieved via the introduction over each strip of the graph
of Boundary Insertion Operators (BIO) 
$\psi^{\lambda(p) \lambda(q)}_{\lambda(p,q)}$ which act as a
coordinate dependent homomorphism from 
$V_{\lambda(p)} \star V_{\lambda(p,q)}$ and $V_{\lambda(q)}$, so
mediating the changing in boundary conditions.
Here $V_{\lambda(\bullet)}$ denotes the Verma module generated by the
action of the Virasoro generators over the $\lambda(\bullet)$ highest
weight and $\star$ denotes the fusion of the two representations.  

In the limit in which the theory is rational (\ie when the
compactification radius is an integer multiple of the self dual
radius $R_{s.d.} \,=\, L(k)/\sqrt{2}$) the compactified boson
theory is the same as an $SU(2)_{k=1}$ WZW 
model, thus it is possible to identify the BIO as primary operators with
well defined conformal dimension and correlators. 
Moreover, considering the
coordinates of three points in
the neighborhood of a generic vertex of the ribbon graph, we can write the OPEs describing the insertion of such operators in
each vertex. Considering four adjacent boundary components, it is then   
possible to show that the
OPE coefficients $C_{j_{(r,p)}j_{(q,r)}j_{(p,q)}}^{j_{p}j_{r}j_{q}}$ 
are provided by the fusion matrices 
$F_{j_{r}j_{(p,q)}}\left[\begin{smallmatrix}
j_p & j_q\\
j_{(r,p)} & j_{(q,r)}\end{smallmatrix}\right]$, 
which in WZW models coincide with the $6j$-symbols of the
quantum group $SU(2)_{e^{\frac{\pi}{3} i}}$:
\begin{equation*}
C_{j_{(r,p)}j_{(q,r)}j_{(p,q)}}^{j_{p}j_{r}j_{q}}=\left\{
\begin{smallmatrix}
j_{(r,p)} & j_{p} & j_{r} \\
j_{q} & j_{(q,r)} & j_{(p,q)}
\end{smallmatrix}
\right\} _{Q=e^{\frac{\pi }{3} i}}
\end{equation*}

From these data, through edge-vertex factorization we can characterize 
the general structure of the partition function for this
model \cite{Mauro2} as a sum over all possible $SU(2)$ primary
quantum numbers describing
the propagation of the Virasoro modes along the $N_0$ cylinders 
$\{\Delta^*_{\varepsilon(k)}\}$. 

The overall picture which emerges 
is that of $N_0$ cylindrical ends glued through their inner boundaries to
the ribbon graph, while their outer boundaries lay on D-branes. Each
D-brane acts naturally as a source for gauge fields: it allows us
to introduce open string degrees of freedom whose information is traded
through the cylinder to the ribbon graph, whose edges thus acquire
naturally a gauge
coloring.
This provides a new kinematical set-up for discussing gauge/gravity
correspondence \cite{Noi1}. 

%%%%%%%%%%%%%%%%%%%%%%%%%%%%%%%%%%%%%%%%%%%%%%%%
%% BACKMATTER
%%%%%%%%%%%%%%%%%%%%%%%%%%%%%%%%%%%%%%%%%%%%%%%%
%
%\begin{theacknowledgments}
%The author wish to tanks 
%\end{theacknowledgments}
%%%%%%%%%%%%%%%%%%%%%%%%%%%%%%%%%%%%%%%%%%%%%%%%
%% You may have to change the BibTeX style below, depending on your
%% setup or preferences.
%%
%% If the bibliography is produced without BibTeX comment out the
%% following lines and see the aipguide.pdf for further information.
%%
%% For The AIP proceedings layouts use either
%%%%%%%%%%%%%%%%%%%%%%%%%%%%%%%%%%%%%%%%%%%%
%\bibliographystyle{unsrt}
\bibliographystyle{aipproc}   % if natbib is available

\begin{thebibliography}{9}
\expandafter\ifx\csname natexlab\endcsname\relax\def\natexlab#1{#1}\fi
\providecommand{\enquote}[1]{``#1''}
\expandafter\ifx\csname url\endcsname\relax
  \def\url#1{\texttt{#1}}\fi
\expandafter\ifx\csname urlprefix\endcsname\relax\def\urlprefix{URL }\fi

\bibitem[Gopakumar and Vafa(1999)]{Gop-Vafa}
Gopakumar, R., and Vafa, C., \emph{Adv. Theor. Math. Phys.}, \textbf{3}, 1415
  (1999), [hep-th/9811131].

\bibitem[Gaiotto and Rastelli(2003)]{Gai-Ras}
Gaiotto, D., and Rastelli, L. (2003), [hep-th/0312196].

\bibitem[Gopakumar(2004{\natexlab{a}})]{Gop1}
Gopakumar, R., \emph{Phys. Rev.}, \textbf{D70}, 025009 (2004{\natexlab{a}}),
  [hep-th/0308184].

\bibitem[Gopakumar(2004{\natexlab{b}})]{Gop2}
Gopakumar, R., \emph{Phys. Rev.}, \textbf{D70}, 025010 (2004{\natexlab{b}}),
  [hep-th/0402063].

\bibitem[Carfora et~al.(2002)]{Mauro1}
Carfora, M., Dappiaggi, C., and Marzuoli, A., \emph{Class. Quant. Grav.},
  \textbf{19}, 5195 (2002), [gr-qc/0206077].

\bibitem[Carfora and Marzuoli(2003)]{Mauro3}
Carfora, M., and Marzuoli, A., \emph{Adv. Theor. Math. Phys.}, \textbf{6},
  357--401 (2003), [math-ph/0107028].

\bibitem[Langlands et~al.(1999)]{Langlands}
Langlands, R.~P., Lewis, M.-A., and Saint-Aubin, Y. (1999), [hep-th/9904088].

\bibitem[Arcioni et~al.(2004)]{Mauro2}
Arcioni, G., Carfora, M., Dappiaggi, C., and Marzuoli, A., \emph{Jour. Geom.
  Phys.}, \textbf{52}, 137 (2004), [hep-th/0209031].

\bibitem[Carfora et~al.(2004)]{Noi1}
Carfora, M., Dappiaggi, C., and Gili, V. (2004), in preparation.

\end{thebibliography}
%\bibliographystyle{aipprocl} % if natbib is missing
%%%%%%%%%%%%%%%%%%%%%%%%%%%%%%%%%%%%%%%%%%%
%% BIBLIOGRAPHY
%%%%%%%%%%%%%%%%%%%%%%%%%%%%%%%%%%%%%%%%%%%

\end{document}